# Software Effort Estimation Accuracy Prediction of Machine Learning Techniques: A Systematic Performance Evaluation

Yasir Mahmood[*,a,b]  , Nazri Kama[a], Azri Azmi[a], Ahmad Salman Khan[b], Mazlan Ali[c]

[a]Advanced Informatics Department, Razak Faculty of Technology and Informatics, Universiti Teknologi Malaysia, 54100 Kuala Lumpur, Malaysia
[b]Department of Software Engineering, Faculty of Information Technology, The University of Lahore, Lahore, Pakistan
[c]Perdana Centre, Razak Faculty of Technology and Informatics, Universiti Teknologi Malaysia, 54100 Kuala Lumpur, Malaysia

*Corresponding author: Yasir Mahmood (mahmood.yasir@graduate.utm.my).

**Abstract:** Software effort estimation accuracy is a key factor in effective planning, controlling and to deliver a successful software project within budget and schedule. The overestimation and underestimation both are the key challenges for future software development, henceforth there is a continuous need for accuracy in software effort estimation (SEE). The researchers and practitioners are striving to identify which machine learning estimation technique gives more accurate results based on evaluation measures, datasets and the other relevant attributes. The authors of related research are generally not aware of previously published results of machine learning effort estimation techniques. The main aim of this study is to assist the researchers to know which machine learning technique yields the promising effort estimation accuracy prediction in the software development. In this paper, the performance of the machine learning ensemble technique is investigated with the solo technique based on two most commonly used accuracy evaluation metrics. We used the systematic literature review methodology proposed by Kitchenham and Charters. This includes searching for the most relevant papers, applying quality assessment criteria, extracting data and drawing results. We have evaluated a state-of-the-art accuracy performance of 28 selected studies (14 ensemble, 14 solo) using Mean Magnitude of Relative Error (MMRE) and PRED (25) as a set of reliable accuracy metrics for performance evaluation of accuracy among two techniques to report the research questions stated in this study. We found that machine learning techniques are the most frequently implemented in the construction of ensemble effort estimation (EEE) techniques. The results of this study revealed that the EEE techniques usually yield a promising estimation accuracy than the solo techniques.

**Keywords:** Software development; software effort estimation; effort estimation accuracy; machine learning; solo methods; ensemble techniques.

## 1. Introduction

In software development perspective, effort estimation is a process of predicting the amount of work and hours required to develop a successful software system on schedule and within the allotted budget plan. It is usually expressed in man-hours or man-months unit. The software effort estimation process is normally done at an early stage of software development where generally a little helpful information is accessible for the estimates. The investigations of software effort estimation have started since the 1960s and have been persistent research because of numerous arguments in accomplishing an accurate software effort estimation results [1,2]. A survey is conducted by Project Management Institute (PMI) in 2017, investigated that 69% software were successfully met the original goals and business intent of the project, 43% were not finished within their initial budgets, 48% were delivered late and 15% failed due to inaccurate effort estimation [3]. In the literature of SEE, there exist several estimation techniques that are classified under three main categories [4]: 1) algorithmic, 2) non-algorithmic and 3) machine learning shown in Figure 1. Algorithmic techniques use statistical and mathematical formulation for software effort estimation. Function Point Analysis (FPA), COCOMO-II (COnstructive COst MOdel), Source Line of Code (SLOC), Putnam Software LIfe cycle Management (SLIM), Use Case Point (UCP) are broadly used estimation techniques. Non-algorithmic models are based on analytical assessments and interpretations. For the analysis, these models use historical data from past finished projects. The expert judgment, planning poker, wideband Delphi and Work Breakdown Structure (WBS) are non-algorithmic techniques. Machine learning techniques are alternatives to algorithmic models. Artificial Neural Networks (ANN), Case-Based Reasoning (CBR), Support Vector Regression (SVR), decision trees, Fuzzy model,Bayesian networks, and Genetic Algorithms (GA) come under the category of machine learning estimation techniques.

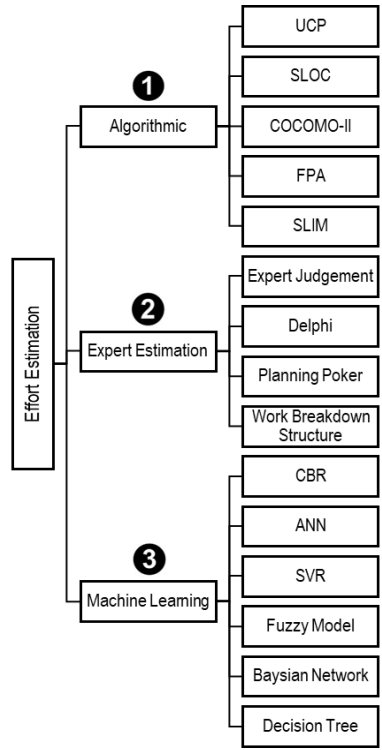

**Figure 1.** Categorization of effort estimation models

In light of the above estimation solo techniques, experts and researchers proposed to make numerous estimation methods and afterwards chosen just a single best method to utilize [5]. Recently, new endeavors on ensemble estimation methods have been proposed [6-8] to avoid the limitations of SEE techniques and merge their favorable advantages. The ensemble effort estimation technique consists of combinations of more than one single technique to estimate software development effort of a new project using a combination rule (mean, median, IRWM, etc.) [9]. The ensemble effort estimation process is shown in Figure 2 summarized by [10]. The combination of the estimates of each base model that creates the ensemble produced the estimation of an ensemble. In the past couple of decades, lots of research have been conducted on various types of effort estimation techniques and proposed lots of models for accomplishing high effort estimation accuracy. However, there is no consensus between the research communities that concludes the best solo method.

Alan Albrecht first developed the Function Point Analysis (FPA) approach in 1979 [47] which was used to measure the size of a software product concerning its features. Several other measurement methods have been developed based on Albrecht's method. International Function Point Users Group (IFPUG) FPA [48], Mark II FPA [49], Netherlands Software Users Metrics Association (NESMA) FPA, Finnish Software Metrics Association (FiSMA), Functional Size Measurement (FSM) and Common Software Measurement International Consortium (COSMIC) are well-known models that are accepted as international standards for FSM by ISO/IEC. Each of these methods of measuring the functionality differs from each other in terms of the metrics and also the rules applied when determining the functional size of the project [50]. Nonetheless, all of these approaches are in particular applicable to information system development projects. The effort estimation based on these measures usually does not longer consider the software development approach used [51]. A few simplifications have been proposed to overcome the problems [52]. In the late 1990's, Abran and Symons suggested COSMIC FP to address certain weaknesses of conventional FPA methodologies such as IFPUG and Mark II [53]. Borandag, E., et al. measured the size of the software project through the function points method and Mark II FPA [54]. In this study, the comparison was made on the average size of ten projects developed by different organizational groups. The results indicated that the FP method estimated a lower error rate compared to MK II FPA for the size of the software projects. In addition, FP and MK II FPA can be used to estimate the size for small and large scale software projects.

The technique proposed by Jodpimai, P., et al. [39] selects only a single method for a GA-based ensemble using a correlation-based feature selection algorithm. The ensemble uses a GA to construct a mathematical function from those selected method estimates to determine one combined estimations. The results indicated that not only the single method but also comparative ensemble methods outperformed the proposed ensemble approach. The experiment was conducted by Hidmi, O., et al [40] to evaluate k-Nearest Neighbor (k-NN) and support vector machine by applying into Maxwell and Desharnais datasets to improve effort accuracy prediction of software development effort. The results of this experiment showed that the effort accuracy equals to 85% when applying a solo method and 91.35% using the combined classifiers. Alhazmi, O.H. and Khan, M.Z. [41] examined the effect of two feature selection algorithms Best Fit and Genetic Algorithm on six ensemble learning algorithms. The Genetic Algorithm (GA) feature selection for the bagging M5Rule was the best method for predicting software development effort. A comparative study on particle swarm-based optimization feature selection for predicting software effort using ensemble methods and machine learning was addressed in [42]. This study focussed on addressing the use of bagging with the said base learners. The results of this study showed the best performance of the bagging M5Rule with an MMRE value of 28.78%. The three machine learning techniques (Logistic Regression, Naïve Bayes and Random Forests.) were applied to a preprocessed COCOMO NASA benchmark data. All the applied techniques were successful in achieving significant results compared to COCOMO model [43].

Silhavy, P., investigated the outliner detection method for software effort estimation models [44]. The accuracy of the estimate was compared with a public and in-house datasets using stepwise regression models. The findings showed that in-house datasets increase the accuracy of estimations. The Median Deviation (MAD) approach provides the optimal estimation accuracy by the use of a stepwise model. It outperforms the FPA method when all criteria are evaluated.

The performance of effort estimation models is highly dependent on the evaluation accuracy metrics which play a key role to determine the effectiveness of that particular model under estimation. Software metrics are techniques to measure some specific characteristics of software products. The term "software metrics" is directly associated with the measurement which is important to achieve the management objectives of prediction, progress, and the process improvement. There are several accuracy measures available to estimate the accurate effort for different SEE models. The MMRE and PRED (25) are based on the magnitude of relative error (MRE) measure. It is defined as the ratio between actual effort and estimated effort.

The MRE is computed by the following equation (1).

$$\text{MRE} = \frac{|\text{actual effort} - \text{estimated effort}|}{\text{actual effort}}$$

(1)

The MRE values are utilized to find the value of MMRE which is determined by the following equation (2).

$$\text{MMRE} = \frac{1}{n}\sum_{1}^{n}\text{MRE}_i$$

*where "n" is the number of observations* (2)

The MMRE is used to demonstrate the relative amount by which the estimated effort is an under-estimate or over-estimates in contrast with the actual estimate. MMRE is utilized in a large portion of the research work as an assessment standard because of its independent-of-units characteristic like person-hours, person-months or man-hours and so on. MMRE is an important instrument used to outline measurements and to assess software effort estimation models. PRED (25) is the percentage of estimates that are within 25% of the actual efforts.

This paper summarizes the performance evaluation of studies concerning effort estimation accuracy in the machine learning-based ensemble and solo techniques. The remainder of this paper is organized as follows: Section 2 represents the related work. Section 3 describes the methodology of selected studies. Section 4 presents the overview of ensemble technique. Section 5 explains the accuracy improvement of the ensemble and solo techniques. Section 6 describes the results and discussion. Threats to validity is described in section 7. The conclusion and future work are described in section 8 and 9 respectively.

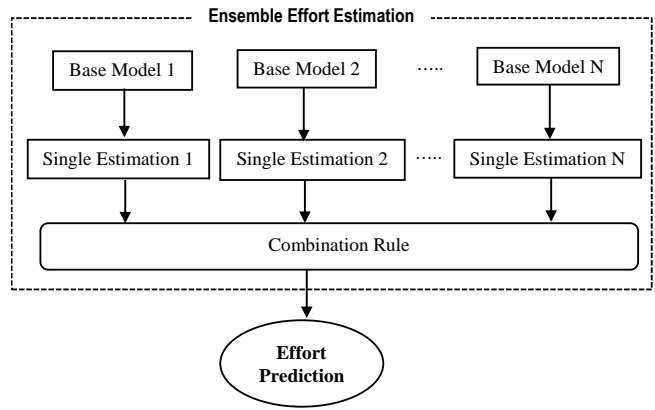

**Figure 2.** Ensemble effort estimation (EEE) process

## 2. Related Work

The success of any software project mainly depends on its effort estimation accuracy. A lot of research has been conducted so far to quantify the accuracy of effort estimation models utilizing distinctive techniques. In any case, researchers and specialists are striving to recognize which estimation technique gives increasingly accurate outcomes on the given datasets and the other applicable attributes. In SEE literature, the researchers have proposed different models and techniques for accomplishing high effort estimation accuracy. Tronto, I. F. d. B., et al. [11] conducted a comparison of artificial neural networks (ANN) and regression models. ANN and regression analysis were applied to COCOMO dataset for estimating effort from size. The performance of both methods was compared and results revealed that ANN was effective in effort estimation. A systematic review was performed by Jorgensen & Shepperd [12] in which 11 estimation methods were being identified. They found that the regression method has been the most frequently used techniques. Wen et al. [13] compared two machine learning techniques and found that it is more accurate than non-machine learning. They investigated that the mean of PRED (25) was 46% and MMRE was 51% for case-based reasoning (CBR) as compared to mean of PRED (25) = 64% and MMRE = 37% for ANN.

Kamal, M. W. and M. A. Ahmed [14] performed a comparison between several UCP metrics and proposed a use case-based framework using fuzzy logic. The results showed that the UCP method may bring a significant impact on the accuracy of estimations in machine learning techniques. Azzeh, M., and A. B. Nassif [15] presented a hybrid model for predicting the effort using Radial Basis Neural Network (RBNN) and Support Vector Machine (SVM), which included classification and prediction stages. In this model, the historical productivity was clustered into fine-grain productivity using bisecting K-medoids algorithm clustering technique and then classified based on environmental factors. The results found that the use of RBNN showed substantial improvement in estimating effort. They also explored the risk of omitting the environmental complexity factor (ECF) from the estimate and the productivity factor should be more focussed. Nagar, C. and A. Dixit [16] merged the UCP and COCOMO models and divided four software projects into sub-modules to estimate the KLOC using use cases. Dividing the project into smaller sub-modules was found to bring the estimated effort closer to the actual effort relative to the entire project.

Gaussian Mixture Model Clustering (GMM), Moving Window (MW), K-means clustering, and Spectral Clustering (SC) techniques were evaluated by Silhavy, R., et al. [17] as a subset selection methods for estimating UCP. It was investigated that the prediction error of linear regression methods due to clustering methods has been significantly reduced. The SC reduced a prediction error by up to 98% when compared to UCP. The moving window produced inconsistent results because of its data sensitivity. Toka, D. and O. Turetken [18] presented an empirical assessment on parametric software estimation models (SLIM, COCOMO II, TruePlanning, and SEER-SEM based on their prediction accuracy. The results suggested that COCOMO-II model showed significant results than the other three models on MMRE metrics. Patil, L. V., et al. [19] used a fuzzy logic technique to improve the accuracy of component-based software development (CBSD) effort estimation and found that component point was the best method for accurate size estimation. Sabrjoo, S., et al. [20] compared the COCOMO model using KLOC with COCOMO using Function Point Analysis (FPA). They investigated that the COCOMO model used with FPA has given more accurate results than KLOC using MMRE evaluation measure.

The COCOMO II model contains three sub-models: (i) Applications Composition, (ii) Early Design, and (iii) Post-Architecture. The application composition model uses productivity rate and function points to estimate the effort and schedule. Additionally, the other sub-models use SLOC as size input and use the same algorithm to measure the estimated effort. However, the early design sub-model should be used while the project is in its early stages and the detailed information about the project is not known. Whereas, post-architecture sub-model will be used when detailed project information such as its designs and environment is available. In a systematic literature review focusing on effort estimation in agile software development, COCOMO was not present among the methods found in the studies included in the review [34]. In a related survey study on effort estimation, 5.26% of the respondents reported that they used COCOMO, but only together with other methods [35]. A case study is presented with the application of COCOMO model and Function Point Analysis (FPA) for the estimation of effort [36]. The case study results for the effort estimation using Function Point Analysis (FPA) indicated the estimate of effort (34.30 man-months), the actual effort (30.12 man-months), and MRE (13.8%). Arnuphaptrairong, T. [37], reported the accuracy of Function Point Analysis (FPA) of 13 software projects using data available in the data flow diagram and showed the MMRE 1624.31%. Kemerer, C. F. [38], analyzed many COCOMO models and investigated that the COCOMO Intermediate model showed the least Mean Magnitude of Relative Error (MMRE=583.82) of the 15 software projects. Ochodek [45] approximated COSMIC and IFPUG FPA functional size using list of use case names or a UML use case diagram. The proposed method was compared with the study of Hussain et al. [46] and concluded a low error prediction.

The researchers claimed that despite of numerous suggested research studies, there is no ultimate consensus on the best solo estimation model. Several models have been proposed to counter the estimation; however, the literature suggests many ways to avoid weaknesses of solo techniques [30]. MacDonell & Shepperd [31,32] suggest combining different solo techniques to form an ensemble; if no foremost technique can be found applied. Similarly, Jørgensen [33] also recommended using more than one model which can hypothetically improve effort estimation accuracy as compared to solo estimation models. To support the literature, this study investigated the software development effort estimation accuracy of solo and ensemble machine learning techniques using two most widely used evaluation metrics to assist the researchers to know which machine learning technique yields the optimal estimation accuracy.

## 3. Methodology

This study is conducted in view of the following activities.

*3.1. Search Strategy*

We found the primary studies through six digital libraries: IEEExplor, ACMDigital Library, Science Direct, Web of Science, SCOPUS and Google Scholar. These digital libraries are most preferred and widely used by the effort estimation community. To find relevant studies based on the research questions shown in Figure 3, the search query string is applied to an individual database.

> software development effort* AND
> (estimation OR prediction OR precision) AND (accuracy OR improvement) AND (techniques OR methods OR models) AND (ensemble OR solo OR single) AND language (English)

**Figure 3.** Search String

Figure. 4 shows the search strategy for the literature review process. A total of 365 research papers was found and downloaded from each database. A library was created in Endnote[1] version X8.2, where the downloaded studies were merged and duplicates removed. Upon eliminating duplicates, 167 studies were eventually chosen. By distributing the selected papers among five authors, the inclusion/exclusion criteria were carried out by manually reviewing the title, abstract, introduction and conclusion. Finally, a consensus was reached among the authors which resulted in 28 research papers out of 365 results passing the quality assessment criteria. Each time a new paper was found, the relevant paper was added to the list.

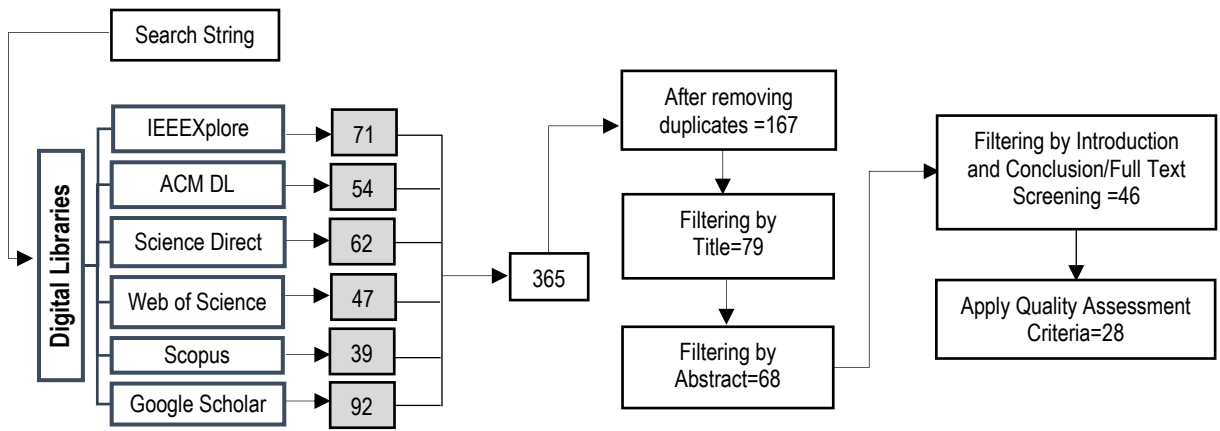

**Figure 4.** Search Strategy

*3.2. Research Questions*

RQ1: What does existing literature reveal about accuracy improvement in ensemble and solo machine learning techniques?

RQ2: Does the ensemble machine learning technique outperforms the solo technique in software effort estimation?

RQ3: Is ensemble technique more accurate in terms of MMRE and PRED (25) on publicly domain (PD) and non-publicly domain (NPD) datasets?

*3.3. Inclusion Criteria*

The studies concerning that are:

a. written in English
b. based on software development ensemble and solo machine learning models, methods or techniques
c. focussed in effort estimation or prediction

---

[1] http://endnote.com

d. based on accuracy improvement or enhancement
   e. published in conference or journal

*3.4. Exclusion Criteria*

a. The studies were excluded that are:
b. not based on MMRE or PRED (25) evaluation measures
c. concerned with effort estimation in software maintenance and testing phase
d. related to non-machine learning techniques
e. concerned with effort distribution

*3.5. Quality Assessment (QA)*

A quality assessment checklist proposed by [28] was customized to qualitatively evaluate the studies. This checklist has also been customized by [29] in their SLRs studies. Table 1 shows the quality assessment checklist used to evaluate our primary studies. A total of 12 questions was customized using the QA checklist for quality measurement, rigorousness, credibility, and relevance of the studies. Every QA checklist question has employed a three-point scale, i.e. Yes (Y) implies 1 point, No (N) is 0 point and Partial (P) is 0.5 points. Thus, the highest score is 12. The quality score of this study is computed by summing up the scores of the answers to the QA questions. We considered only the related studies with an acceptable quality score greater than 6 (50% of perfect score), meaning that if any selected study scored 6 or below will be eliminated from our list of studies.

**Table 1.** Quality assessment checklist

| Questions | Score |
|---|---|
| 1. Are the research aims clearly specified? | Y/N/P |
| 2. Was the study designed to achieve these aims? | Y/N/P |
| 3. Are the estimation techniques used clearly described and their selection justified? | Y/N/P |
| 4. Are the variables considered by the study suitably measured? | Y/N/P |
| 5. Are the data collection methods adequately described? | Y/N/P |
| 6. Is the data collected adequately described? | Y/N/P |
| 7. Is the purpose of the data analysis clear? | Y/N/P |
| 8. Are statistical techniques used to analyze data adequately described and their use justified? | Y/N/P |
| 9. Do the researchers discuss any problems with the validity/reliability of their results? | Y/N/P |
| 10. Are all research questions answered adequately? | Y/N/P |
| 11. Are the study findings credible? | Y/N/P |
| 12. How clear are the links between data, interpretation and conclusions? | Y/N/P |

In view of this, we have divided our primary studies among the authors (Author_1=9, Author_2=9, Author_3=9, Author_4=10, Author_5=9). The five authors of this review performed the quality assessment of the relevant studies independently. All disagreements on the results of the quality assessment were discussed among all authors, and eventually, consensus was reached. As a result, a total of 18 papers was failed to achieve the inclusion score of above 6. Table 2 and Figure 5 summarize the quality scores of the remaining 28 primary studies.

Table 2. Quality score of primary studies

| ID | QA1 | QA2 | QA3 | QA4 | QA5 | QA6 | QA7 | QA8 | QA9 | QA10 | QA11 | QA12 | Score |
|---|---|---|---|---|---|---|---|---|---|---|---|---|---|
| S1 | 1 | 1 | 0.5 | 1 | 1 | 0 | 1 | 0 | 0 | 0 | 1 | 0.5 | 7 |
| S2 | 1 | 1 | 1 | 1 | 1 | 0 | 1 | 1 | 0.5 | 0 | 1 | 0.5 | 9 |
| S3 | 0.5 | 1 | 1 | 1 | 0.5 | 1 | 0 | 1 | 0 | 0 | 0.5 | 1 | 7.5 |
| S4 | 1 | 1 | 0.5 | 0 | 1 | 1 | 0 | 0 | 0 | 0.5 | 1 | 1 | 7 |
| S5 | 1 | 1 | 1 | 0.5 | 0 | 1 | 1 | 0 | 1 | 1 | 0 | 0 | 7.5 |
| S6 | 1 | 0.5 | 1 | 0 | 1 | 1 | 0.5 | 1 | 0 | 1 | 0 | 1 | 8 |
| S7 | 0 | 1 | 1 | 1 | 0.5 | 0.5 | 0 | 1 | 0.5 | 0.5 | 0.5 | 0.5 | 7 |
| S8 | 1 | 1 | 1 | 0 | 1 | 1 | 1 | 0 | 1 | 1 | 0.5 | 0.5 | 9 |
| S9 | 0 | 1 | 1 | 1 | 1 | 1 | 1 | 1 | 0.5 | 0 | 1 | 0 | 8.5 |
| S10 | 1 | 1 | 0.5 | 1 | 1 | 1 | 0 | 1 | 0 | 1 | 1 | 1 | 9.5 |
| S11 | 1 | 0.5 | 0.5 | 1 | 0 | 0 | 1 | 0 | 1 | 1 | 0 | 1 | 7 |
| S12 | 1 | 1 | 0 | 0 | 1 | 1 | 0 | 0 | 0.5 | 0.5 | 1 | 0 | 6 |
| S13 | 0.5 | 0 | 1 | 0.5 | 1 | 0 | 0.5 | 0 | 1 | 1 | 0 | 0 | 6.5 |
| S14 | 1 | 0.5 | 1 | 0 | 0 | 1 | 1 | 1 | 0 | 1 | 0.5 | 0 | 7 |
| S15 | 1 | 0.5 | 0 | 1 | 1 | 1 | 1 | 1 | 1 | 0 | 0.5 | 1 | 9 |
| S16 | 0.5 | 1 | 1 | 1 | 0 | 0 | 1 | 1 | 0 | 1 | 0.5 | 0 | 7 |
| S17 | 1 | 0 | 0 | 1 | 1 | 1 | 1 | 0 | 0 | 0 | 1 | 0.5 | 6.5 |
| S18 | 1 | 1 | 1 | 1 | 0 | 1 | 1 | 1 | 0.5 | 1 | 1 | 1 | 10.5 |
| S19 | 1 | 1 | 0 | 0 | 1 | 0.5 | 1 | 1 | 0 | 1 | 1 | 1 | 8.5 |
| S20 | 1 | 0 | 1 | 1 | 1 | 1 | 1 | 1 | 0 | 0 | 1 | 1 | 9 |
| S21 | 0.5 | 1 | 1 | 1 | 0.5 | 1 | 0.5 | 0.5 | 0 | 0 | 0.5 | 0 | 6.5 |
| S22 | 1 | 1 | 0 | 0.5 | 1 | 0.5 | 1 | 1 | 1 | 1 | 1 | 1 | 10 |
| S23 | 1 | 0.5 | 1 | 0.5 | 1 | 1 | 1 | 1 | 0.5 | 0.5 | 0.5 | 1 | 9.5 |
| S24 | 1 | 0 | 0 | 0.5 | 1 | 1 | 1 | 1 | 1 | 0.5 | 0 | 1 | 8 |
| S25 | 0.5 | 0.5 | 1 | 1 | 0 | 0 | 1 | 1 | 0.5 | 1 | 0 | 1 | 7.5 |
| S26 | 0.5 | 1 | 0 | 1 | 1 | 0 | 1 | 0 | 1 | 0.5 | 0 | 1 | 7 |
| S27 | 0 | 1 | 0 | 0 | 1 | 1 | 1 | 0.5 | 1 | 1 | 1 | 1 | 8.5 |
| S28 | 1 | 0 | 0 | 0 | 1 | 1 | 0 | 0 | 1 | 1 | 1 | 0.5 | 6.5 |
| Total | 22.0 | 20.0 | 17.0 | 17.5 | 20.5 | 19.5 | 20.5 | 17.0 | 13.5 | 17.0 | 17.0 | 18.0 | 220.5 |
| Average | 0.79 | 0.71 | 0.61 | 0.63 | 0.73 | 0.70 | 0.73 | 0.61 | 0.48 | 0.61 | 0.61 | 0.64 | 7.88 |

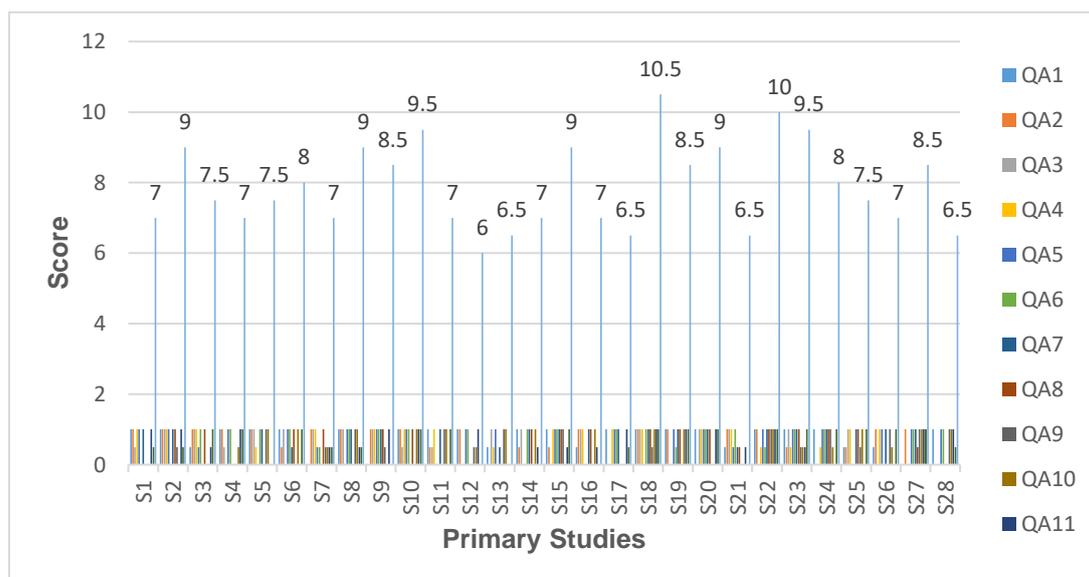

Figure 5. Quality score

## 4. Outline of an Ensemble Technique and Accuracy Evaluation Metrics

The thought behind utilizing EEE in SEE is that each single estimation technique has its qualities and shortcomings by consolidating techniques through EEE. We can relieve the shortcomings, which may lead to more accurate estimation that can be acquired from single models [21,22].

There are two types of EEE methods [23]:

1- Homogeneous EEE: used to refer to an ensemble that consolidates one base model with no less than two distinct combinations of one ensemble learning.

2- Heterogeneous EEE: used to refer to an ensemble that incorporates two diverse base models or more.

The researchers have conducted various empirical studies to evaluate ensemble effort estimation techniques and some of them have focused on dealing only with homogeneous ensembles, heterogeneous ensembles, or both types of EEE techniques [24]. The literature stated that the base methods that compose the ensemble ought to fulfill two conditions: high accuracy and diversity to achieve a high estimation accuracy. In other words, the base methods ought to be diverse and as accurate as could be expected under the circumstances. Thus, each base technique can cancel the estimation errors done by other base methods. Otherwise, an ensemble that integrates no diverse base methods may produce a lower estimation accuracy than its base methods [25]. The architecture of ensemble is shown in Figure 6 [26], where "X" represents the feature vector for a project under estimation. The same set of feature vectors is provided to all the other predictive algorithms (P1, P2... Pn) and provides estimations (y1, y2... yn). Based on the estimations provided by the individual predictive algorithm, the ensemble aggregator "f" aggregate estimation using combination rules (mean, median, IRWM by weights w1, w2...wn). In the end, it provides the ensemble estimation for the project ($Y_{ensemble}$).

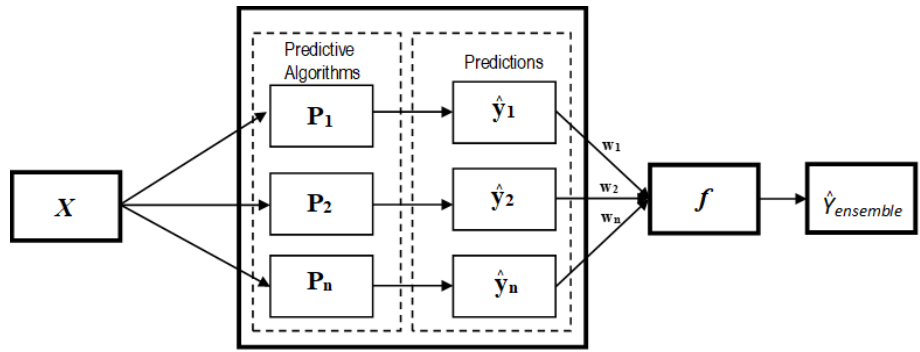

**Figure 6.** Ensemble-based architecture

In order to evaluate the estimation models, appropriate metrics are needed to measure their accuracy. There is a number of accuracy measures available to evaluate the prediction of estimation results for different SEE models. The study investigated by Mahmood et al. [29], showed that the mean magnitude of relative error (MMRE) and the percentage of prediction (PRED 25) have been most commonly used in total 10 (S2, S6, S7, S10, S11, S12, S18, S21, S22, S23) and 8 (S2, S5, S6, S7, S16, S21, S22, S23) studies out of 25 respectively. Specifically, MRE, MSE, MMER and RMSE are used in 4 (S6, S7, S16, S17), 4 (S2, S4, S10, S14), 4 (S5, S10, S11, S22) and 3 (S4, S10, S14) respectively. Finally 28% (7 studies) did not report the accuracy metrics used to validate the models. The detail of accuracy metrics used to evaluate the estimation models is shown in Table 3 [29].

## 5. Accuracy Performance of Primary Studies

Table 4 and Table 5 report the performance accuracy of ensemble and solo techniques based on machine learning in estimating software development effort. A summarized contribution of the selected studies along with the estimation technique is stated. It should be noted that in our study, we have taken accuracy values under best configuration among different datasets and techniques so that accurate performance analysis may be investigated.

**TABLE 3:** Accuracy evaluation metrics [29]

| Cited study | MRE | MMRE | MdMR | PRED (.10) | PRED (.25) | PRED (.50) | PRED (.75) | PRED (X) | AE | SA | MAE | MBRE | MIBRE | MSE | MMER | MAPE | RMSE | NRMS | SSE | SD | R² | RSS | AICc | PRE/ AMSE | NS[a] |
|---|---|---|---|---|---|---|---|---|---|---|---|---|---|---|---|---|---|---|---|---|---|---|---|---|---|
| [S1]  |   |   |   |   |   |   |   |   |   |   |   |   |   |   |   |   |   |   |   |   |   |   |   |   | ✓ |
| [S2]  |   | ✓ | ✓ |   | ✓ | ✓ |   |   |   |   |   |   |   | ✓ |   |   |   |   |   |   |   |   |   |   |   |
| [S3]  |   |   |   |   |   |   |   |   | ✓ | ✓ | ✓ | ✓ | ✓ |   |   |   |   |   |   |   |   |   |   |   |   |
| [S4]  |   |   |   |   |   |   |   |   |   |   |   |   |   | ✓ |   | ✓ | ✓ | ✓ | ✓ |   |   |   |   |   |   |
| [S5]  |   |   |   | ✓ | ✓ | ✓ | ✓ |   |   |   |   |   |   |   | ✓ |   |   |   |   |   |   |   |   |   |   |
| [S6]  | ✓ | ✓ | ✓ | ✓ | ✓ |   |   |   |   |   |   |   |   |   |   |   |   |   |   |   |   |   |   |   |   |
| [S7]  | ✓ | ✓ |   |   | ✓ |   |   |   |   |   |   |   |   |   |   |   |   |   |   |   |   |   |   |   |   |
| [S8]  |   |   |   |   |   |   |   |   |   |   |   |   |   |   |   |   |   |   |   |   |   |   |   |   | ✓ |
| [S9]  |   |   |   |   |   |   |   |   |   |   |   |   |   |   |   |   |   |   |   |   |   |   |   |   | ✓ |
| [S10] |   | ✓ |   |   |   |   |   | ✓ |   |   |   |   |   | ✓ | ✓ |   | ✓ |   |   |   |   |   |   |   |   |
| [S11] |   | ✓ |   |   |   |   |   |   |   |   |   |   |   |   | ✓ |   |   |   |   | ✓ |   |   |   |   |   |
| [S12] |   | ✓ |   |   |   |   |   | ✓ |   |   |   |   |   |   |   |   |   |   |   |   |   |   |   |   |   |
| [S13] |   |   |   |   |   |   |   |   |   |   |   |   |   |   |   |   |   |   |   |   |   |   |   |   | ✓ |
| [S14] |   |   |   |   |   |   |   |   |   |   |   |   |   | ✓ |   |   | ✓ |   |   |   | ✓ | ✓ | ✓ |   |   |
| [S15] |   |   |   |   |   |   |   |   |   |   |   |   |   |   |   |   |   |   |   |   |   |   |   |   | ✓ |
| [S16] | ✓ |   |   |   | ✓ |   |   |   |   |   |   |   |   |   |   |   |   |   |   |   |   |   |   |   |   |
| [S17] | ✓ |   |   |   |   |   |   |   |   |   |   |   |   |   |   |   |   |   |   |   |   |   |   |   |   |
| [S18] |   | ✓ |   |   |   |   |   |   |   |   |   |   |   |   |   |   |   |   |   |   |   |   |   |   |   |
| [S19] |   |   |   |   |   |   |   |   |   |   |   |   |   |   |   |   |   |   |   |   |   |   |   | ✓ |   |
| [S20] | difference of > 5 % of the average of efforts estimated by UCP and COCOMO[b] |||||||||||||||||||||||| |
| [S21] |   | ✓ |   |   | ✓ |   |   |   |   |   |   |   |   |   |   |   |   |   |   |   |   |   |   | ✓ |   |
| [S22] |   | ✓ |   |   | ✓ |   |   |   |   |   |   |   |   |   | ✓ |   |   |   |   |   | ✓ |   | ✓ |   |   |
| [S23] |   | ✓ |   |   | ✓ |   |   |   |   |   |   |   |   |   |   |   |   |   |   |   |   |   |   |   |   |
| [S24] |   |   |   |   |   |   |   |   |   |   |   |   |   |   |   |   |   |   |   |   |   |   |   |   | ✓ |
| [S25] |   |   |   |   |   |   |   |   |   |   |   |   |   |   |   |   |   |   |   |   |   |   |   |   | ✓ |
| Count | 4 | 10 | 2 | 2 | 8 | 2 | 1 | 2 | 1 | 1 | 1 | 1 | 1 | 4 | 4 | 1 | 3 | 1 | 1 | 1 | 2 | 1 | 2 | 2 | 7 |
| %     | 16 | 40 | 8 | 8 | 32 | 8 | 4 | 8 | 4 | 4 | 4 | 4 | 4 | 16 | 16 | 4 | 12 | 4 | 4 | 4 | 8 | 4 | 8 | 8 | 28 |

[a]Not Specified. [b]Other.

*5.1 Accuracy of Solo Technique*

Nassif, A. B., et al. [S1] proposed a Treeboost model which was evaluated against the use case point (UCP) model and multiple linear regression model using 58 industrial and 26 education projects. The result of this model indicated that the treeboost model outperformed the regression and the UCP model by 9% and 15% based on the PRED (25) and MMRE criteria respectively. The MMRE value of treeboost model was 0.29 and PRED (25) was 64% which indicated its significance. To improve the estimation of software effort, Nassif, A. B., et al. [S2] applied a Sugeno fuzzy inference system to a regression model. This model showed significant improvement accuracy of 11% over Karner's model and 7% over Schneider's model based on MMRE evaluation measure.

Table 4. Prediction accuracy of primary studies concerning solo techniques

| | Cited study | Author(s) Ref. | Estimation technique/contribution | Dataset | Prediction accuracy | |
|---|---|---|---|---|---|---|
| | | | | | MMRE | PRED (25) |
| Dataset Availability: Non-Public Domain (NPD) | [S1] | Nassif, A. B., et al. | Treeboost (Stochastic Gradient Boosting) model to predict software effort based on use case points method. | Dataset-1: 58 industrial projects Dataset-2: 26 educational projects | 0.29 | 64 |
| | [S3] | Satapathy, S. M., et al. | Random forest (RF) was used to improve the prediction values and assessed the software effort using UCP. | 149 software projects | 0.33 | 68.45 |
| | [S5] | Malhotra, R. and A. Jain | Evaluated and compared Linear Regression, ANN, DT, SVM and Bagging on software project dataset. | 499 software projects | 0.17[a] | 52[a] |
| | [S7] | Yurdakurban, V. and N. ErdoĞan | Decision Tree, Naive Bayes and Multiple Regression models were inspected. | Data obtained from a local software house | 0.15[a] | 83[a] |
| | [S8] | Sharma, P. and J. Singh | Employed Random Forest (RF), Multilayer Perceptron (MLP) and Support Vector Machines (SVM). | 4 software projects | 0.30[a] | 72.09[a] |
| | [S11] | Nassif, A. B., Capretz, L. F., and Ho, D. | Proposed a new feed-forward Artificial Neural Network (ANN) model to predict software effort based on the use case points model. | Dataset-1: 214 industrial projects Dataset-2: 26 educational projects | 0.49 | 29.16 |
| | [S14] | S. M. Satapathy, B. P. Acharya, and S. K. Rath | Stochastic Gradient Boosting (SGB) technique is used to obtain better prediction accuracy with class point approach. | 40 software projects | 0.43 | 95.32 |
| Dataset Availability: Public Domain (PD) | [S2] | Nassif, A. B., et al. | A regression model was proposed based on the UCP model. A Sugeno Fuzzy Inference System (FIS) approach was applied to this model to improve the estimation. | Comparison with Karner [18] and Schneider models | 0.28 | 54 |
| | [S4] | Dan, Z. | Artificial neural network (ANN) model incorporated with COCOMO which was improved by applying particle swarm optimization (PSO). | COCOMO I NASA | 0.40[a] | 55.10[a] |
| | [S6] | Baskeles, B., et al. | Developed a learning model using machine learning techniques MLP, RBF, DT and SVM. | NASA, USC and SDR | 0.12 | 88.33 |
| | [S9] | Idri, A., et al. | Integrates the fuzzy k-prototypes technique into the fuzzy analogy estimation process. | ISBSG, COCOMO, USP05-FT, and USP05-FT | 0.52[a] | 56.76[a] |
| | [S10] | Rijwani, P. and S. Jain | Proposed the usage of ANN using Multi-Layered Feed Forward Neural Network. | COCOMO | 0.14 | 43[a] |
| | [S12] | Gabrani, G., & Saini, N. | The comparative study of various machine learning techniques used for estimating the software effort by empirical evaluation of five different evolutionary learning algorithms | Desharnais, Maxwell, Miyazaki | 0.36[a] | 33[a] |
| | [S13] | Burgess, C. J., & Lefley, M. | Critically evaluated the potential of Genetic Programming (GP) in software effort estimation. | Desharnais | 0.37 | 28 |

[a] Accuracy value under best configuration among datasets/techniques
[b] Mean among software projects

Moreover, the fuzzy logic approach improved PRED (25) by 16.5% over Karnar's model. Random forest (RF), a machine learning technique was used by Satapathy, S. M., et al. [S3] to improve the prediction values and assessed the software effort using UCP approach and obtained promising results. The results revealed that the MMRE = 0.33 and PRED (25) = 68.45% using Random Forest (RF) fuzzy approach outperformed the SGB, LLR, RBFN and MLP approach. The MMRE value of 0.33 surpasses MLP, RBFN, SGB and LLR by 34.69%, 70.89%, 25.43% and 5.22% respectively. Similarly, the PRED (25) value of 68.45% outclassed these techniques by 30.20%, 32.88%, 8.72% and 31.35%. Artificial Neural Network (ANN) model was incorporated with COCOMO by applying particle swarm optimization by Dan, Z. [S4]. The PSO-ANNCOCOMO II showed an improvement of 3.27% than the original ANN-COCOMO II. It should be noted that for MMRE evaluation measure, the smaller value is better and for PRED (25) the larger value is better. The MMRE value of PSO-ANN-COCOMO II was 0.40 which was smaller as compared to MMRE = 0.42 of ANN-COCOMO II model on NASA dataset.

The PSO-ANN-COCOMO II showed 3.1% improvement (PRED (25) = 55.10%) over ANN-COCOMO II (PRED (25) = 52%). Hence, the effort estimation accuracy of PSO-ANN-COCOMO II model was better than PSO-ANN-COCOMO II model. Malhotra, R. and A. Jain [S5] compared and evaluated machine learning techniques and proved that the Decision Tree (DT) was better in accuracy with MMRE = 0.17 and PRED (25) = 52% than ANN, SVM, Linear Regression and Bagging methods. An experiment has been conducted by Baskeles, B., et al. [S6] using MLP, RBF, DT and SVM machine learning techniques on three datasets. The results of this study found that RBF on USC datasets outperformed other techniques with MMRE = 0.12 and PRED (25) = 88.33%. Yurdakurban, V. and N. ErdoĞan [S7] inspected three machine learning techniques out of those regression analysis with MMRE = 0.15 and PRED (25) = 83% surpassed Naive Bayes and Multiple Regression models. The regression techniques showed 14% and 9% more accurate estimation against DT and Naive Bayes techniques respectively. Similarly, for PRED (25) 33% and 16% improvement are achieved. The UCP effort size metric was used by Sharma, P. and J. Singh [S8] conducted an experiment on Support Vector Machines (SVM), Multilayer Perceptron (MLP) and Random Forest (RF) for the accurate software effort estimation. The experimental results obtained from RF were better as compared to MLP and SVM. The 17.3% and 26.49% for MMRE; 4.65% and 25.58% for PRED (25) improvement are achieved against MLP and SVM techniques respectively. A novel integrated approach was proposed by Idri, A., et al. [S9] to handle numerical and categorical attributes merged with fuzzy k-prototypes technique into fuzzy analogy estimation process over four datasets through Jackknife evaluation. The 2FA-kmodes results of MMRE = 0.52 and PRED (25) = 56.76% using USP05_FT were better than other techniques over remaining datasets. The usage of ANN (Artificial Neural Network) using Multi-Layered Feed Forward Neural Network was proposed by Rijwani, P. and S. Jain [S10] and obtained a significant MMRE value of 0.14 as compared to a MMRE value of 0.36 with COCOMO.

Nassif, A. B., Capretz, L. F., and Ho, D. [S11], proposed ANN model to predict software effort based on the use case point model. The proposed ANN model, based on MMRE, outperforms the Regression and UCP models by 8% and 50% respectively. The UCP model marginally exceeds the ANN model based on PRED (0.25). Gabrani, G., & Saini, N. [S12], concluded that the evolutionary algorithms have better results for estimating software effort than other approaches. Burgess, C. J., & Lefley, M. [S13], evaluated Genetic Programming (GP) in the software effort estimation. The findings showed that the GP is significantly more accurate in terms of MMRE but rather poor results for PRED (25). S. M. Satapathy, B. P. Acharya, and S. K. Rath [S14], used Stochastic Gradient Boosting (SGB) technique to obtain better prediction accuracy with class point approach. The results suggested that the effort estimation model of Stochastic Gradient Boosting (SGB-based) has lower MMRE, NRMSE, and higher prediction accuracy. Therefore, it can be concluded that the effort estimation using the SGB-based model performs more accurately than the MLP and RBFN-based models.

*5.2 Accuracy of Ensemble Technique*

A lot of research has been conducted so far on ensemble software development estimation. A correlation-based feature selection algorithm was utilized to choose an important single technique for genetic algorithm-based ensemble [S15]. The several single estimation strategies were being ensemble which delivered distinctive estimations however comparably high accuracy. The ensemble method used in this approach produced high estimation accuracy with MMRE = 0.16 on Desharnais dataset and PRED (25) = 70% with Telecom dataset. A heterogeneous EEE was investigated by Hosni, M., et al. [S16] based on four machine learning techniques using three linear rules and three datasets. The ensemble of M5P, MLP, k-nn and SVR used Inverse Ranked Weighted Mean (IRWM) combination rule on KEMERER dataset has achieved high-performance accuracy of 60% in terms of PRED (25).

Table 5. Prediction accuracy of primary studies concerning ensemble techniques

| Cited study | | Author(s) Ref. | Estimation technique/contribution | Dataset | Prediction accuracy | |
|---|---|---|---|---|---|---|
| | | | | | MMRE | PRED (25) |
| Dataset Availability: Non-Public Domain (NPD) | [S17] | Kumar, A. and U. Datta | Adaptive Neuro-Fuzzy Inference model was considered with bagging and boosting to predict future values. | 11 industrial projects | 0.12[b] | 69.37[b] |
| | [S24] | Minku, L. L. and X. Yao | Evaluated ensemble methods over single learning machines. MLP, Bagging, RBF, NCL | Cross-company Org 1-7 | 0.10[a] | 71.70[a] |
| | [S25] | Pospieszny, P., et al | Proposed an ensemble averaging of three machine learning algorithms; Support Vector Machines, Neural Networks and Generalized Linear Models. | 11 variables of software projects | 0.17 | 69.44 |
| | [S26] | Azzeh M., et al | Presented two approaches to measure the similarity between two software projects based on fuzzy C-means clustering and fuzzy logic. | 5 projects | 0.13 | 84 |
| | [S27] | Shepperd, M., & Schofield, C. | Described a technique based upon the use of Case-Based Reasoning (CBR) with stepwise regression analysis using mean rule. | 21 real-time projects | 0.23 | 74 |
| | [S28] | Seo, Y.-S., et al | Applied Least square and K-means clustering, and three effort estimation methods; Least squares, Neural network and Bayesian network. | Bank data from single financial company | 0.27[a] | 70[a] |
| Dataset Availability: Public Domain (PD) | [S15] | Jodpimai, P., et al. | A correlation-based feature selection algorithm was used to select only the necessary single method for the genetic algorithm-based ensemble. | Desharnais, COCOMO81, Maxwell, Miyazaki94, Telecom Kemerer | 0.16[a] | 70[a] |
| | [S16] | Hosni, M., et al. | Heterogeneous ensemble based on K-NN, MLP, SVR and M5P solo techniques using three combiner rules. | Albrecht, Kemerer, Miyazaki | 0.13 | 60[a] |
| | [S18] | Wu, D., et al. | Six case-based reasoning (CBR) methods (Euc-CBR Man-CBR, Min-CBR, Gre-CBR, Gau-CBR, Mah-CBR) with optimized weights derived from the particle swarm optimization (PSO) method. | Desharnais, Miyazaki | 0.32[a] | 51[a] |
| | [S19] | Elish, M. O. | Voting ensemble model was compared over five solo models MLP, RBF, RT, KNN and SVR. | Albrecht, Miyazaki, Maxwell, COCOMO, Desharnais | 0.47[a] | 37.50[a] |
| | [S20] | Hsu, C., et al. | Integrated COCOMO, regression, analogy, GRA and ANN to make an ensemble over seven datasets using five combination rules. | Albrecht, Bailey, COCOMO, Desharnais, Jørgensen A, Jørgensen B, Kemerer | 0.26 | 72.22[a] |
| | [S21] | Kultur, Y., et al. | Proposed an ensemble neural network algorithm (ENNA) using associative memory by MLP. | NASA, NASA93, USC, SDR, Desharnais | 0.10[a] | 72.22[a] |
| | [S22] | Li, Q., et al. | Introduced the systematic "external" combining idea estimated software effort using Optimal Linear Combining (OLC) method. | F. Kemerer's [27] | 0.10 | 83.37 |
| | [S23] | Braga, P. L., et al. | Investigated the use of bagging predictors applied on M5P/regression trees, MLP, linear regression and SVR. | NASA | 0.16[a] | 88.89[a] |

[a] Accuracy value under best configuration among datasets/techniques
[b] Mean among software projects

Kumar, A. and U. Datta [S17] proposed an ensemble learning methods to predict future value with the help of Adaptive Neuro-Fuzzy Inference model with bagging and boosting. It was observed that the Neuro-Fuzzy model produced better results than all other models in term of PRED (25) = 0.21. Six case-based reasoning (CBR) methods (Euc-CBR, Man-CBR, Min-CBR, Gre-CBR, Gau-CBR, Mah-CBR) derived from the particle swarm optimization (PSO) method were proposed by [S18]. They have achieved PRED (25) = 51% and MMRE = 0.32 values with Miyazaki dataset using WMC combination rule. The highest improvement was 6.78% at the training stage and 20.34% at the testing stage of MMRE evaluation criteria. For the PRED (25), the highest improvement was 4.33% at the training stage and 2.55% at the testing stage. The five solo models MLP, RBF, RT, KNN and SVR were compared using voting ensemble model [S19]. The results confirmed that the solo models are not reliable in terms of accuracy as they perform differently across different datasets. The MMRE value of 0.61 and PRED (25) value of 37.50% for ensemble model outperformed all other the solo models.

Hsu, C., et al. [S20] integrated COCOMO, regression, analogy, GRA and ANN to make an ensemble over seven datasets using five combination rules. The proposed model showed experimental results and showed that this model was useful for improving estimation accuracy. The ensemble uses Bailey dataset, the MMRE = 0.26 and PRED (25) = 72.22% surpassed solo methods in all other datasets.

The stable and accurate estimations are attainable through neural networks using ensemble with associative memory Kultur, Y., et al. [S21]. The experimental results MMRE = 0.01 and PRED (25) = 72.22% showed that the proposed algorithm (ENNA) produced better results than the neural networks (NN) in terms of accuracy and robustness. Li, Q., et al. [S22] introduced the systematic external combining idea estimated software effort using Optimal Linear Combining (OLC) method. The MMRE = 0.10 and PRED (25) = 83.37 indicated its significance. The use of bagging predictors applied to M5P/regression trees, MLP, linear regression and SVR were investigated by Braga, P. L., et al. [S23]. The results revealed that bagging has improved the performance of M5P, regression, MLP and SVR with MMRE value of 0.16 and PRED (25) value of 88.89%. Minku, L. L. and X. Yao [S24] evaluated the available ensemble methods to improve SEE given by single learning machines and investigated its usefulness among other methods. The ensemble methods outperformed based on MMRE = 0.10 and PRED (25) = 71.70% values. Pospieszny, P., Czarnacka-Chrobot, B., & Kobylinski, A. [S25], proposed an ensemble averaging of three machine learning algorithms; Support Vector Machines, Neural Networks and Generalized Linear Models. The ensemble models were slightly less accurate than SVM, however, the effort ensemble model characterized MMRE below 0.2 and PRED (0.25) at a 70% level.

Azzeh M., Neagu D., Cowling P. [26], presented two approaches for measuring the similarity between two fuzzy C-means clustering and fuzzy logic-based software projects. In general, the results showed that the ensemble of three analogies was better than the second approach. Shepperd, M., & Schofield, C. [S27], used Case-Based Reasoning (CBR) with stepwise regression analysis using mean rule. The accuracy level of effort estimation showed promising results with MMRE=0.23 and PRED (25) =74%. Seo, Y.-S., et al [S28], investigated the influence of outlier elimination upon the accuracy of software effort estimation through empirical studies using two outlier elimination methods (Least square and K-means clustering) and three effort estimation methods (Least squares, Neural network and Bayesian network) associatively.

## 6. Results and Discussion

This section reports on the results of the prediction accuracy of the techniques discussed in Section 5. The comparison of solo and ensemble techniques was evaluated using IBM SPSS version 23 statistical analysis tool. The descriptive statistics of an ensemble and solo techniques on Publicly Domain (PD) and Non-Publicly Domain (NPD) datasets based on the values studied in Table 4 and 5 respectively are shown in Table 6. It should be noted that for MMRE evaluation measure, the smaller value is better and for PRED (25) the larger value is better. The mean value of MMRE = 0.17 for an ensemble technique was smaller than the mean value of MMRE = 0.30 for a solo technique on NPD dataset. An ensemble technique showed 13% improvement over solo technique indicated that the ensemble technique was better in accuracy than the solo technique in terms of the mean value of MMRE. Similarly, the mean value of MMRE = 0.21 for an ensemble technique was smaller than the mean value of MMRE = 0.31 for a solo technique on PD dataset and showed 10% improvement over solo technique. Moreover, the mean values of ensemble_MMRE_PD (0.21) and ensemble_MMRE_NPD (0.17) were smaller than the mean values of solo_MMRE_NPD (0.30) and solo_MMRE_PD (0.31) respectively. The results of the mean value of MMRE evaluation measure indicated that the ensemble technique was better in accuracy on both PD and NPD datasets over solo technique.

**Table 6.** Descriptive statistics of ensemble and solo techniques on publicly and non-publicly datasets

| | | solo_MMRE_NPD | solo_MMRE_PD | ensemble_MMRE_NPD | ensemble_MMRE_PD | solo_PRED25_NPD | solo_PRED25_PD | ensemble_PRED25_NPD | ensemble_PRED25_PD |
|---|---|---|---|---|---|---|---|---|---|
| N | Valid | 7 | 7 | 6 | 8 | 7 | 7 | 6 | 8 |
| | Missing | 1 | 1 | 2 | 0 | 1 | 1 | 2 | 0 |
| Mean | | .30 | .31 | **.17** | **.21** | 66.28 | 51.17 | **73.08** | **66.90** |
| Median | | .30 | .36 | **.15** | **.16** | 68.45 | 54.00 | **70.85** | **71.11** |
| Minimum | | .15 | .12 | **.10** | **.10** | 29.16 | 28.00 | **69.37** | **37.50** |
| Maximum | | .49 | .52 | **.27** | **.47** | 95.32 | 88.33 | 84.00 | **88.89** |

The median value of an ensemble using MMRE is 0.16, smaller than solo MMRE = 0.36 on PD dataset, indicated 20% improvement over solo technique. The median value of MMRE = 0.15 for an ensemble technique was smaller than MMRE = 0.30 for a solo technique on NPD dataset showed 15% improvement accuracy over solo technique. The results of the median value of MMRE evaluation measure indicated that the ensemble technique was better in accuracy on both PD and NPD datasets over solo technique. The minimum value of MMRE = 0.10 and maximum of 0.47 achieved by an ensemble technique as compared to solo technique (min = 0.12, max = 0.52) means that an ensemble produced better estimation accuracy with an equitable combination of techniques and combination rules. The graphical representation of MMRE for an ensemble and solo techniques on PD and NPD datasets is shown in Figure 7.

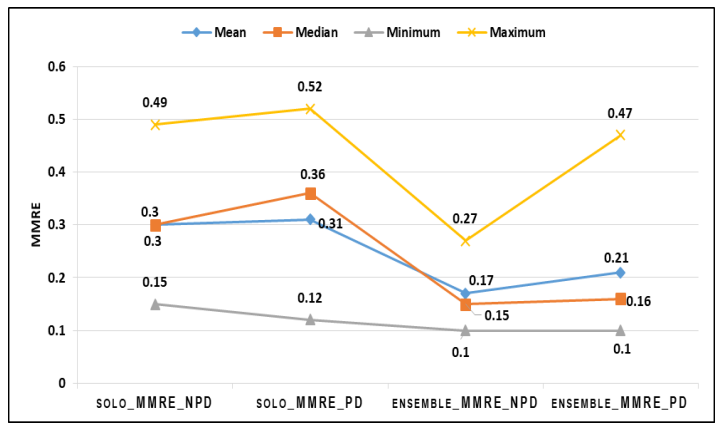

Figure 7. MMRE accuracy of an ensemble and solo techniques on PD and NPD datasets

The mean value of PRED (25) = 66.90% for an ensemble technique was greater than the mean value PRED (25) = 51.17% for a solo technique on PD dataset. An ensemble technique exhibited

15.73% improvement over solo technique. The mean value of PRED (25) = 73.08% on NPD dataset for an ensemble technique was greater than the PRED (25) = 66.28% for a solo technique. An ensemble technique showed 6.80% improvement over solo technique. Moreover, the mean values of ensemble_PRED25_PD (66.90%) and ensemble_PRED25_NPD (73.08%) were greater than the mean values of solo_PRED25_NPD (66.28) and solo_PRED25_PD (51.17%) respectively. The results of the mean value of PRED (25) evaluation measure indicated that ensemble technique was better in accuracy on both PD and NPD datasets over solo technique. Similarly, the median value of an ensemble PRED (25) was 71.11%, which was greater than the value of 54%, showed the improvement of 17.11% over solo technique on PD dataset. Moreover, the median values of ensemble_PRED25_PD (71.11%) and ensemble_PRED25_NPD (70.85%) were greater than the median values of solo_PRED25_NPD (68.45%) and solo_PRED25_PD (54%) respectively. The results of the median value of PRED (25) evaluation metrics indicated that the ensemble technique was better in accuracy on both PD and NPD datasets over solo technique. The maximum value that has been achieved by the ensemble was 88.89% as compared to the solo value of 88.83% on PD dataset. However, the maximum accuracy achieved by the solo technique (max = 95.32%) was better than the ensemble technique (max = 84.00%) on NPD dataset. It should also be noted that the minimum values achieved by ensemble technique, 37.50% and 69.37% on both PD and NPD datasets were better in accuracy than solo technique respectively. The graphical representation of PRED (25) for an ensemble and solo techniques on PD and NPD datasets are shown in Figure 8.

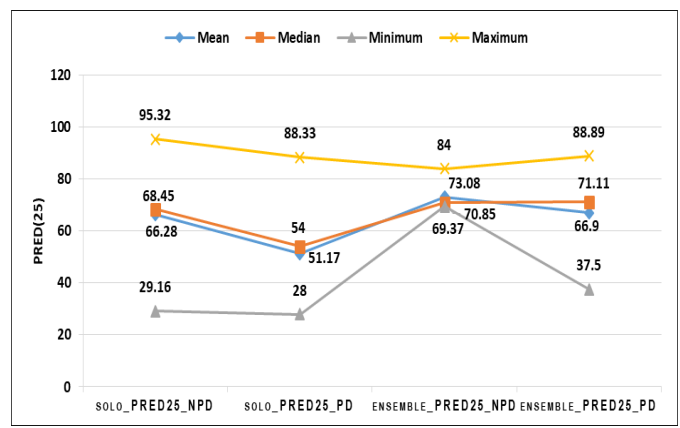

Figure 8. PRED (25) accuracy of an ensemble and solo techniques on PD and NPD datasets

The estimation accuracy of solo and ensemble techniques using PRED (25) evaluation metrics of individual selected studies on NPD and PD datasets is shown in Figure 9 and 10 respectively. There were total 7 studies of solo technique ([S1], [S3], [S5], [S7], [S8], [S11] and [S14]) and 6 studies of ensemble technique ([S17], [S24], [S25], [S26], [S27] and [S28]) for NPD dataset. There were 7 studies of solo technique ([S2], [S4], [S6], [S9], [S10], [S12] and [S13]) and 8 studies of ensemble technique ([S15], [S22], [S23], [S20], [S18], [S16], [S19] and [S21]) for PD dataset.

The ensemble studies [S17], [S24], [S25], [S26], [S27] and [S28] have shown 5.37%, 7.7%, 5.44%, 20%, 10% and 6%, 0.92%, 3.25%, 0.99%, 15.55%, 5.55% and 1.55%, 17.37%, 19.7%, 17.44%, 32%, 22% and 18%, 40.21%, 42.52%, 40.28%, 84.84%, 44.84% and 40.84% accuracy improvement over solo technique in studies [S1], [S3], [S5] and [S11] on NPD dataset. The solo technique of study [S7] and [S8] are better in accuracy in terms of PRED (25) over ensemble technique of studies [S17], [S24] and [S25]. The ensemble studies [S15], [S22], [S23], [S20] and [S16] have shown 16%, 29.37%, 34.89%, 18.22% and 6%, 14.9%, 28.27%, 33.79%, 17.12% and 4.9%, 13.24%, 26.61, 32.13, 15.46 and 3.24%, 27%, 40.37%, 45.89%, 29.22% and 17%, 37%, 50.37%, 55.89%, 39.22 and 27%, 42%, 55.37%, 60.89%, 44.22% and 32% accuracy improvement using PRED (25) over solo technique in studies [S2], [S4], [S9], [S10], [S12] and [S13] on PD dataset.

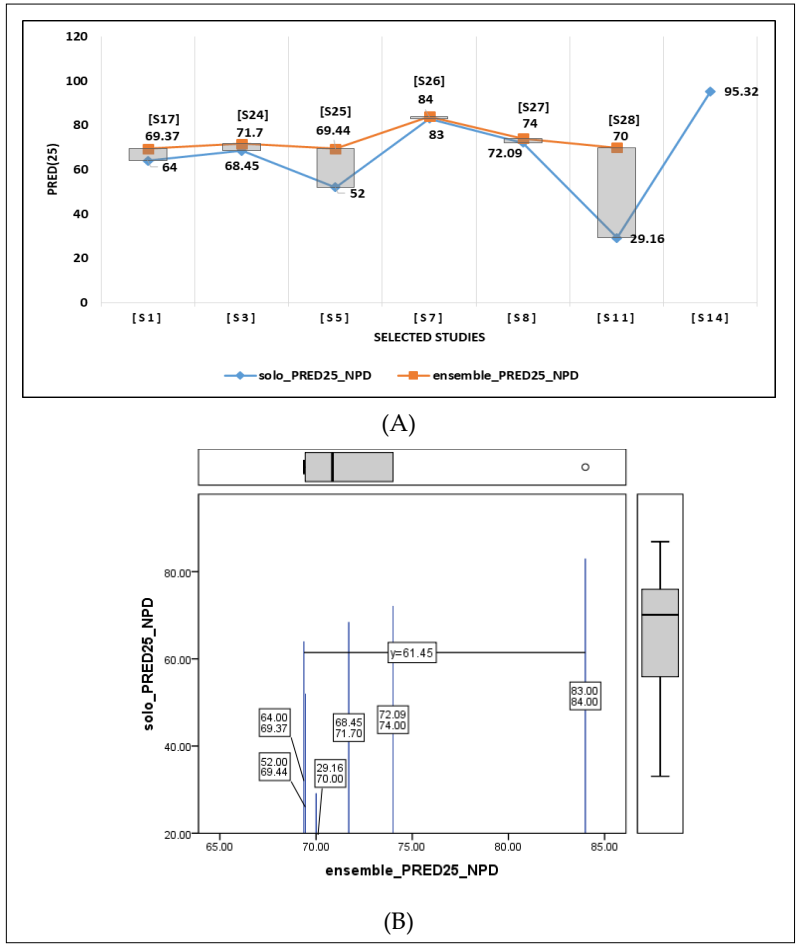

**Figure 9.** PRED (25) accuracy of an ensemble and solo techniques on NPD datasets

The ensemble studies [S18] and [S19] have shown improvement over [S10 for S18 only], [S12] and [S13] solo technique of studies. The solo study [S6] was better in accuracy compared to all ensemble studies except [S23]. The ensemble technique outperformed the solo technique over maximum selected studies on PD and NPD datasets using PRED (25) evaluation measure.

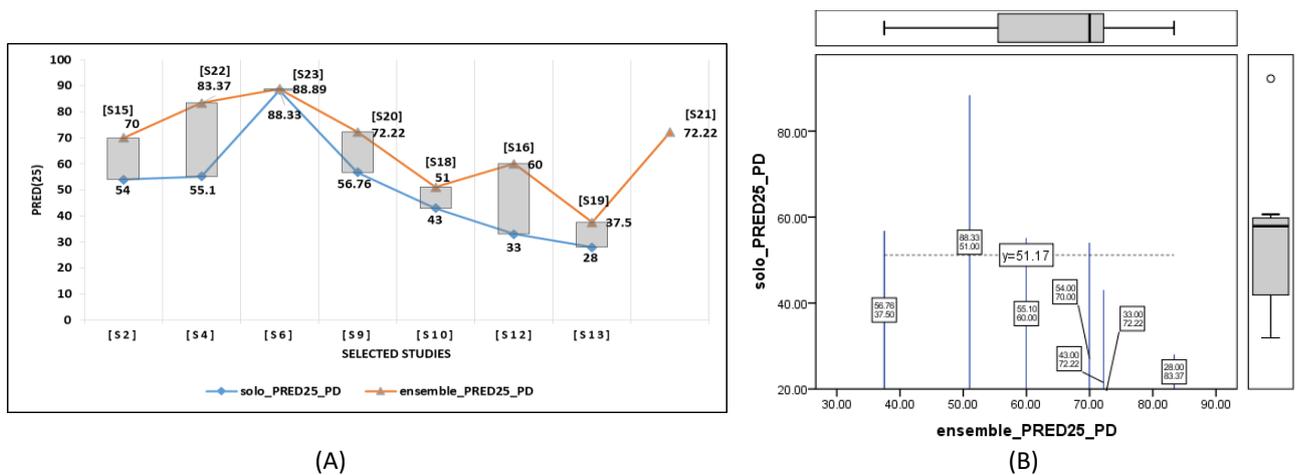

**Figure 10.** PRED (25) accuracy of an ensemble and solo techniques on PD datasets

The estimation accuracy of solo and ensemble techniques using MMRE evaluation metrics of individual selected studies on NPD and PD datasets is shown in Figure 11 and 12 respectively.

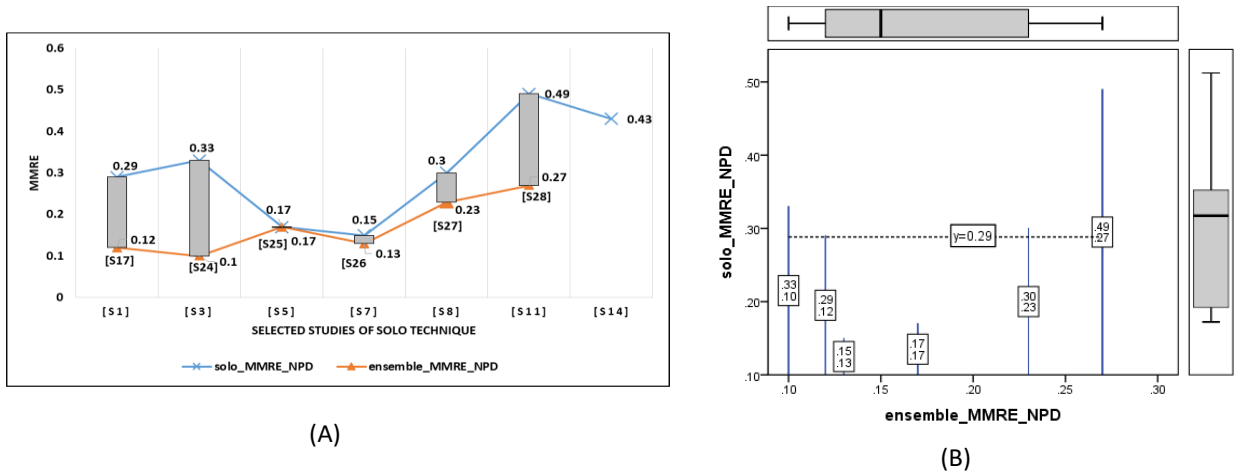

(A)   (B)

**Figure 11.** MMRE accuracy of an ensemble and solo techniques on NPD datasets

The MMRE values of 0.12, 0.1, 0.17 and 0.13 of ensemble studies [S17], [S24], [S25] and [S26] respectively have outperformed the accuracy improvement over solo technique of studies [S1], [S3], [S5], [S7], [S8], [S11] and [S14] on NPD dataset. The ensemble studies [S27] and [S28] have shown accuracy improvement over all solo techniques except [S25] and [S26]. The MMRE value of 0.16 of ensemble studies [S15] and [S23] have shown improvement over all solo technique of studies except [S6] and [S10] on PD dataset. The ensemble studies [S27] and [S28] have shown accuracy improvement over all solo techniques except solo studies [S25] and [S26]. Similarly, the MMRE values of ensemble studies [S16-0.13], [S21-0.1], [S20-0.26] and [S22-0.1] were less than the values of all solo studies except [S6-0.12] for ensemble study [S16-0.13], and [S6-0.12 and S10-0.14] for ensemble [S20-0.26]. The results indicated that the ensemble technique was better in accuracy in terms of MMRE evaluation metrics on PD dataset.

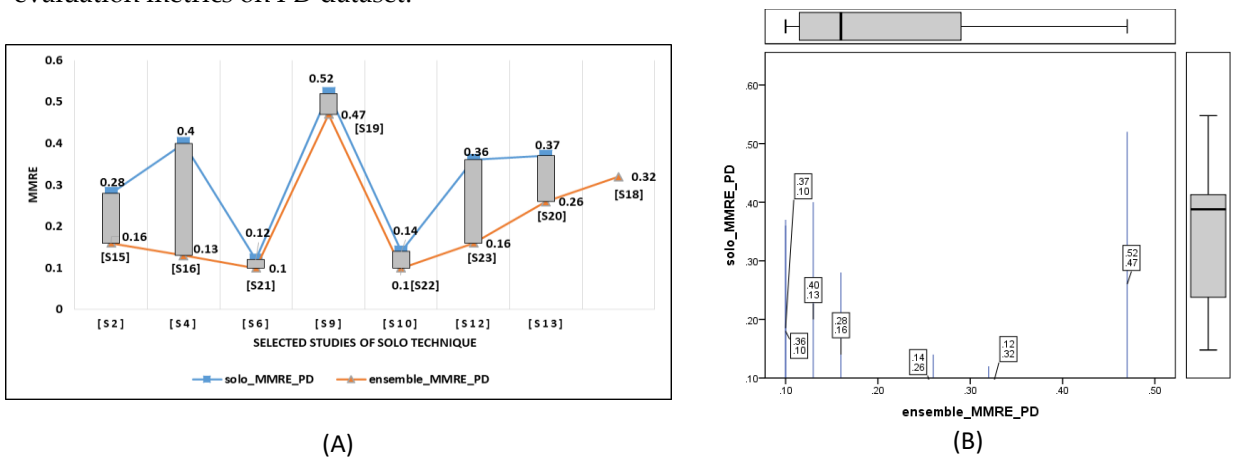

(A)   (B)

**Figure 12.** MMRE accuracy of an ensemble and solo techniques on PD datasets

The MMRE values investigated in our twenty eight selected studies were being arranged in ascending order to assign a rank among the two techniques over PD and NPD datasets. This has been implemented to know which technique has performed better accordingly. The top 10 ranks are taken for the analysis based on MMRE and PRED (25) evaluation measures on both NPD and PD datasets shown in Table 7 and Figure 13.

**Table 7.** Accuracy rankings of ensemble and solo techniques

| MMRE | | | | | | | | PRED(25) | | | | | | | |
|---|---|---|---|---|---|---|---|---|---|---|---|---|---|---|---|
| NPD | | | | PD | | | | NPD | | | | PD | | | |
| Rank | ID | MMRE | Tech | Rank | ID | MMRE | Tech | Rank | ID | PRED(25)% | Tech | Rank | ID | PRED(25)% | Tech |
| 1 | [S24] | 0.1 | E | 1 | [S21] | 0.1 | E | 1 | [S14] | 95.32 | S | 1 | [S23] | 88.89 | E |
| 2 | [S17] | 0.12 | E | 2 | [S22] | 0.1 | E | 2 | [S26] | 84 | E | 2 | [S6] | 88.33 | S |
| 3 | [S26] | 0.13 | E | 3 | [S6] | 0.12 | S | 3 | [S7] | 83 | S | 3 | [S22] | 83.37 | E |
| 4 | [S7] | 0.15 | S | 4 | [S16] | 0.13 | E | 4 | [S27] | 74 | E | 4 | [S20] | 72.22 | E |
| 5 | [S25] | 0.17 | E | 5 | [S10] | 0.14 | S | 5 | [S8] | 72.09 | S | 5 | [S21] | 72.22 | E |
| 6 | [S5] | 0.17 | S | 6 | [S15] | 0.16 | E | 6 | [S24] | 71.7 | E | 6 | [S15] | 70 | E |
| 7 | [S27] | 0.23 | E | 7 | [S23] | 0.16 | E | 7 | [S28] | 70 | E | 7 | [S16] | 60 | E |
| 8 | [S28] | 0.27 | E | 8 | [S20] | 0.26 | E | 8 | [S25] | 69.44 | E | 8 | [S9] | 56.76 | S |
| 9 | [S1] | 0.29 | S | 9 | [S2] | 0.28 | S | 9 | [S17] | 69.37 | E | 9 | [S4] | 55.1 | S |
| 10 | [S8] | 0.3 | S | 10 | [S18] | 0.32 | E | 10 | [S3] | 68.45 | S | 10 | [S2] | 54 | S |
| 11 | [S3] | 0.33 | S | 11 | [S12] | 0.36 | S | 11 | [S1] | 64 | S | 11 | [S18] | 51 | E |
| 12 | [S14] | 0.43 | S | 12 | [S13] | 0.37 | S | 12 | [S5] | 52 | S | 12 | [S10] | 43 | S |
| 13 | [S11] | 0.49 | S | 13 | [S4] | 0.4 | S | 13 | [S11] | 29.16 | S | 13 | [S19] | 37.5 | E |
| | | | | 14 | [S19] | 0.47 | E | | | | | 14 | [S12] | 33 | S |
| | | | | 15 | [S9] | 0.52 | S | | | | | 15 | [S13] | 28 | S |

**Abbreviations:** E, Ensemble; S, Solo;

It was observed that the studies conducted by researchers using ensemble techniques were ranked well than the solo techniques. An ensemble technique was ranked 1, 2, 3, 5, 7, 8, and 1, 2, 4, 6, 7, 8, 10 specified 6 and 7 count numbers as compared to 4 and 3 count numbers (ranked 4, 5, 9, 10, and 3, 5, 9) of a solo technique on NPD and PD datasets.

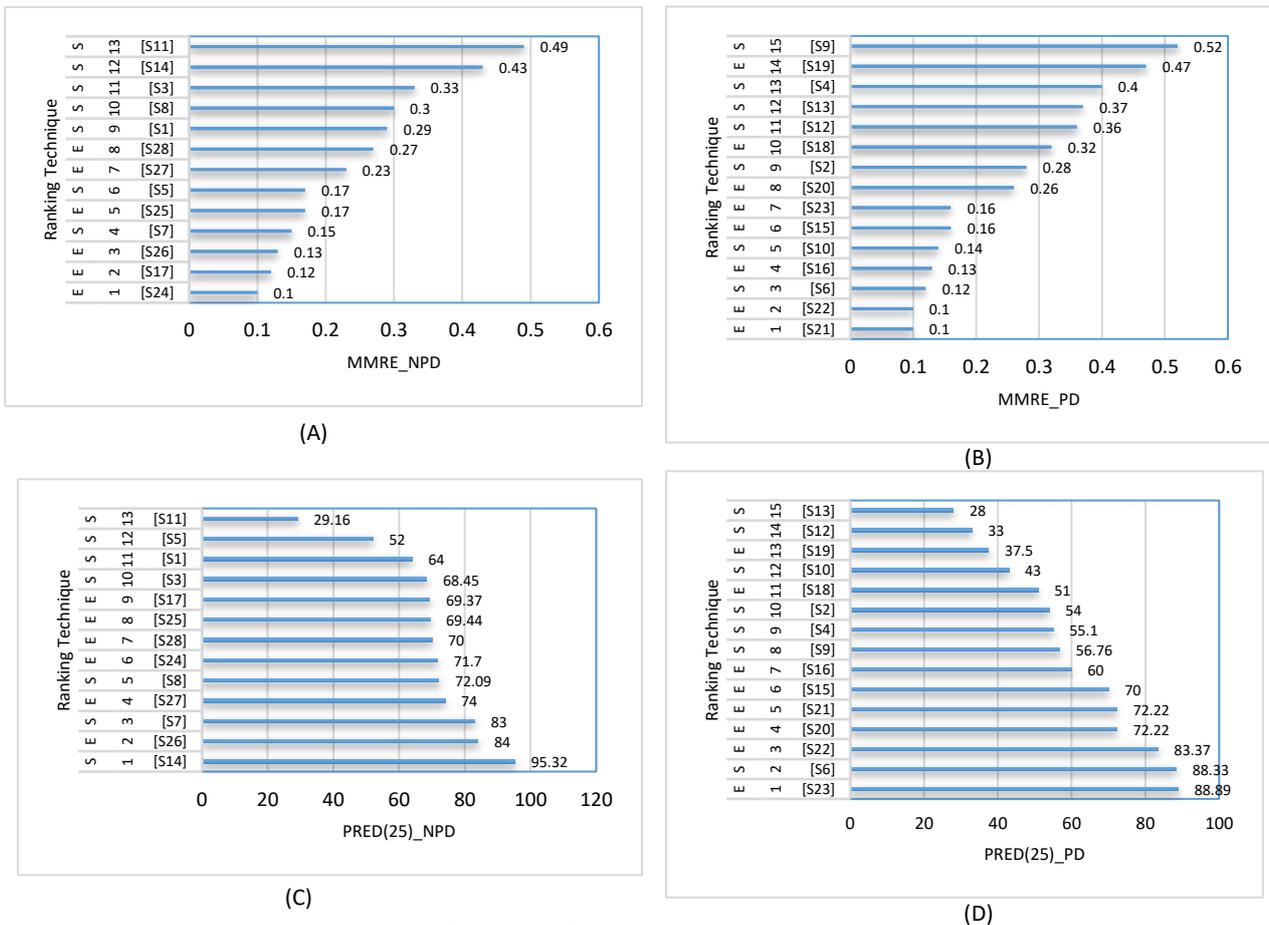

**Figure 13.** Ranking of primary studies

An ensemble technique lies 60% and 70% between MMRE values ranging from 0.1 – 0.3 and 0.1 – 0.32 respectively. The average MMRE value of solo and ensemble technique is 0.31 and 0.17 on NPD dataset, 0.31 and 0.21 on PD dataset indicated 14% and 10% accuracy improvement over solo technique respectively. Using PRED (25), an ensemble technique was ranked 2, 4, 6, 7, 8, 9, and 1, 3, 4, 5, 6, 7 among top ten ranks specified 6 count numbers as compared to 4 count numbers (ranked 1, 3, 5, 10 and 2, 8, 9, 10) of a solo technique on NPD and PD datasets. An ensemble technique lies 60% between PRED (25) values ranging from 68.45% – 95.32% and 54% – 88.89% on both datasets. The average PRED (25) value of solo and ensemble technique is 66.29% and 73.09% on NPD dataset, 51.17% and 66.90% on PD dataset indicated 6.8% and 15.73% accuracy improvement over solo technique. The comparison of the number of counts of accuracy ranking of ensemble and solo techniques in MMRE and PRED (25) evaluation metrics is shown in Figure 14.

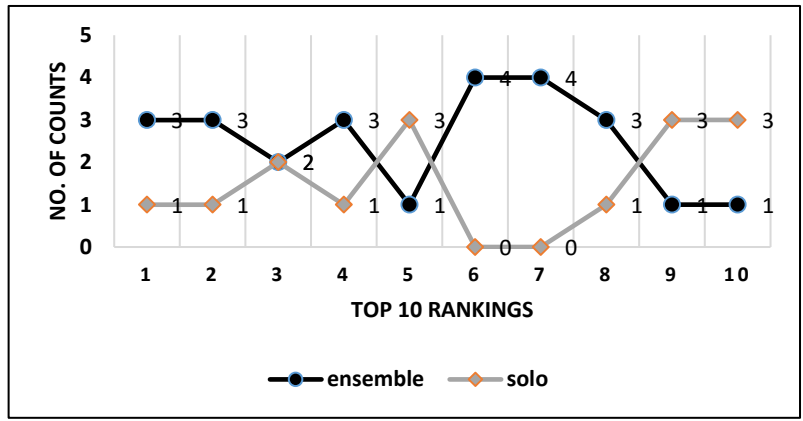

**Figure 14**. Top 10 rankings of an ensemble and solo techniques of MMRE and PRED (25)

In top 10 rankings, the ensemble technique was ranked three times 1, 2, 4, 8, and four times ranked 6 and 7. The solo technique was ranked three times 5, 9 and 10. The total number of counts (20) of ensemble technique was greater compared to solo technique (9) indicated that the ensemble technique outperformed the solo technique in the top 10 rankings. Additionally, in the first step, the nonparametric Wilcoxon Signed-Rank test [55] is performed to compare the performance of two techniques on a single data set. This test compares the ranks for the positive and negative differences in performance of two models/techniques, and is defined in equation (3).

$$\min \sum_{di>0} R(di) + \frac{1}{2} \sum_{di=0} R(di), \sum_{di<0} R(di) + \frac{1}{2} \sum_{di=0} R(di),$$

*where, R(di) is the rank of the difference in performance between two techniques, ignoring signs.*   (3)

Table 8 shows the results of performance ranking comparison of ensemble and solo techniques using Wilcoxon Signed-Rank test. The ensemble technique significantly outperformed the solo technique by sum of ranks on both MMRE and PRED (25) evaluation measures. The MMRE and PRED (25) accuracy performance of an ensemble technique on NPD and PD datasets is significantly ranked higher (mean rank and sum of ranks) than the solo technique. The sum of ranks of an ensemble technique using MMRE metrics on NPD and PD datasets (15.00 and 21.50) are higher than the solo technique (.00 and 3.25). Additionally, the sum of ranks of an ensemble technique using PRED (25) on both the datasets (21.00 and 20.00) are higher than the solo technique (.00 and 8.00).

In the second step of this procedure, the Friedman test [56] is performed, which is a nonparametric version of the repeated measures ANalysis Of VAriance (ANOVA) test. The MMRE and PRED (25) performance of an ensemble and solo techniques are ranked on NPD and PD datasets shown in Figure 15.

Table 8. Performance ranking comparison of ensemble and solo techniques

| Technique on Single Dataset | | N | Mean Rank | Sum of Ranks |
|---|---|---|---|---|
| ensemble_MMRE_NPD solo_MMRE_NPD | Negative Ranks | 5[a] | 3.00 | 15.00 |
| | Positive Ranks | 0[b] | .00 | .00 |
| | Ties | 1[c] | | |
| | Total | 6 | | |
| ensemble_MMRE_PD solo_MMRE_PD | Negative Ranks | 5[d] | 4.30 | 21.50 |
| | Positive Ranks | 2[e] | 3.25 | 6.50 |
| | Ties | 0[f] | | |
| | Total | 7 | | |
| ensemble_PRED25_NPD solo_PRED25_NPD | Negative Ranks | 0[g] | .00 | .00 |
| | Positive Ranks | 6[h] | 3.50 | 21.00 |
| | Ties | 0[i] | | |
| | Total | 6 | | |
| ensemble_PRED25_PD solo_PRED25_PD | Negative Ranks | 2[j] | 4.00 | 8.00 |
| | Positive Ranks | 5[k] | 4.00 | 20.00 |
| | Ties | 0[l] | | |
| | Total | 7 | | |

**Abbreviations:** *a. ensemble_MMRE_NPD < solo_MMRE_NPD, b. ensemble_MMRE_NPD > solo_MMRE_NPD, c. ensemble_MMRE_NPD = solo_MMRE_NPD, d. ensemble_MMRE_PD < solo_MMRE_PD, e. ensemble_MMRE_PD > solo_MMRE_PD, f. ensemble_MMRE_PD = solo_MMRE_PD,*
*g. ensemble_PRED25_NPD < solo_PRED25_NPD. h. ensemble_PRED25_NPD > solo_PRED25_NPD, i. ensemble_PRED25_NPD = solo_PRED25_NPD, j. ensemble_PRED25_PD < solo_PRED25_PD, k. ensemble_PRED25_PD > solo_PRED25_PD, l. ensemble_PRED25_PD = solo_PRED25_PD*

In MMRE evaluation metrics, the rank 1 indicating the best performance and rank 4 the worst. Additionally, in case of PRED (25), the rank 4 indicating the best performance and rank 1 the worst.

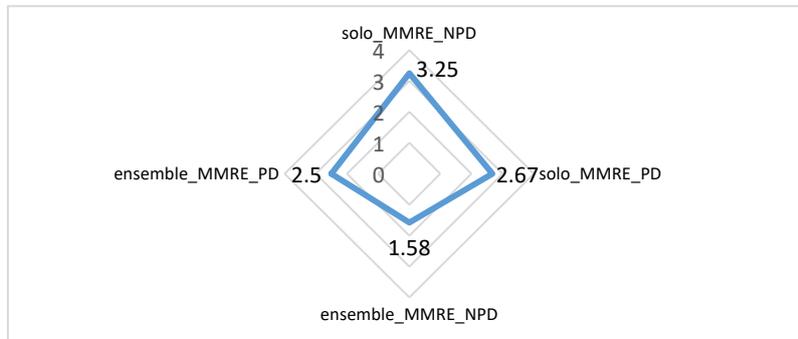

(A)

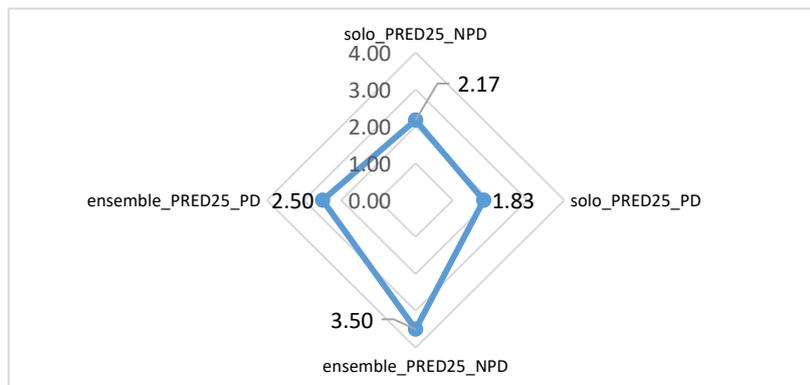

(B)

**Figure 15.** Mean ranking of an ensemble and solo techniques

It can be seen from Figure 15 that the mean rank of an ensemble technique (1.58 and 2.5) is the overall best performing technique using MMRE metrics on both datasets. Similarly, an ensemble technique significantly outperformed the solo technique on both datasets using PRED (25) with a mean rank of 3.50 and 2.50.

**7. Threats to Validity**

This section explains threats to the validity of this paper regarding internal, external and construct validity.

Internal validity: To correctly extract the data from the selected studies, five authors of this study read each paper independently. The data extracted for each paper were compared. All disagreements on data quality were discussed among all authors, and the consensus was eventually reached. The only best estimation accuracy obtained from the optimal configuration of the ensemble and solo techniques was extracted. However, there might still be a bias in data extraction, nevertheless, using the optimal configuration is a good way to minimize the bias. As a result, a total of 28 primary studies listed in Table 2 have been chosen.

External validity: The accuracy values of the estimation were extracted from papers focusing on solo and ensemble effort estimation (EEE) models. In addition, these values were derived from various experimental designs involving the selection of techniques, data sets, data preparation, and employing the method used to validate the model. Note that all of these steps directly affect the accuracy of estimations. It is nonetheless believed that the estimation accuracy results obtained from various experimental designs are more robust than those from a uniform design. This study used publicly and non-publicly domain datasets. The projects are collected from different countries and organizations, and their features are diverse. This makes them adequate for evaluating the accuracy of solo and ensemble effort estimation techniques. However, it will be beneficial to replicate this study using other machine learning techniques and datasets with similar characteristics.

Construct Validity: The purpose of construct validity is to answer the question of the reliability of the performances measured through this study. Since this study focuses only on the accuracy of effort estimates, two evaluation criteria (MMRE, PRED (25)) were used. The main reasons behind this choice are that these performance criteria are unbiased, less fragile, and commonly used in asymmetry assumptions [29]. Additionally, there are several other metrics that are less vulnerable to asymmetry assumptions such as Mean Inverted Balanced Relative Error (MIBRE) and Mean Balanced Relative Error (MBRE) that can also be used to assess the proposed technique.

**8. Conclusion**

In this paper, we contributed to the domain of software effort estimation. We specifically focused on the machine learning-based ensemble and solo techniques for effort estimation. In this research work, our contribution is two-fold. Firstly, we explored the state of the art in the domain of effort estimation using ensemble and solo techniques. We elicited primary studies by following well established Systematic Literature Review (SLR) protocol prescribed for the software engineering domain. Secondly, we compared and evaluated both techniques by applying commonly used accuracy performance metrics (MMRE and PRED (25)) on PD and NPD datasets. The research community may take our investigation as a foundation for further exploration in this domain. Our research work will facilitate the practitioners for deciding on a suitable effort estimation technique for their future software development projects. The main findings of this study to answer our research questions are as follows:

RQ1: We broadly investigated the accuracy improvement of the machine learning ensemble and solo techniques used in software effort estimation. The summarized study to answer this question is also highlighted in Table 4 and Table 5. We found that both ensemble and solo techniques have been utilized for effort estimation in the scientific literature. However, we discovered that the machine learning ensemble techniques perform better for achieving accurate effort estimation results. The main reason behind the better performance of ensemble technique is that unlike solo technique, it utilizes a suitable combination of rules and techniques to predict the effort estimation.

RQ2: This question investigated that the machine learning ensemble technique produced better estimation accuracy results than a solo technique. Our results showed that the ensemble technique outperforms solo technique when evaluated under MMRE and PRED (25) evaluation metrics. The ensemble technique was more accurate on both PD and NPD datasets over solo technique. This is because each solo estimation technique has merits and demerits which leads to somehow inaccurate estimation results. The analysis of this question has been described in section VI.

RQ3: In effort estimation literature, MRE, MMRE and PRED (25) were the most frequently used evaluation measures. The results revealed that an ensemble technique was accurate in terms of MMRE and PRED (25) in most of the selected studies. All our statistical measures showed that the ensemble technique performs better as compared to solo technique. For example, our result showed that an ensemble technique showed 13% and 10% improvement in terms of the mean value of MMRE over solo technique on NPD and PD datasets. Similarly, based on MMRE, an ensemble technique showed 20% and 15% improvement in terms of median over solo technique on both datasets. Moreover, an ensemble technique exhibited 15.73% and 6.80%, 17.11% and 2.40% improvement in terms of mean and median of PRED (25) evaluation measure over solo technique on both datasets. An ensemble technique lies 60% and 70% between MMRE values ranging from 0.1 – 0.3 and 0.1 – 0.32 respectively. The MMRE value of solo and ensemble technique showed 14% and 10% accuracy improvement over solo technique on NPD and PD datasets. An ensemble technique lies 60% between PRED (25) values showed 6.8% and 15.73% accuracy improvement over solo technique on both datasets. Overall, in the top 10 rankings, the ensemble technique was ranked three times 1, 2, 4, 8, and four times ranked 6 and 7. The total number of counts (20) of ensemble technique was greater compared to solo technique (9) indicated that the ensemble technique outperformed the solo technique in the top 10 rankings. The detailed breakdown of this question has been elaborated in section VI.

## 9. Future Work

The investigation of this paper offers different perspectives for future research. The effect of cost factors on the accuracy of the effort estimation models proposed in agile software development will be an interesting research direction. The development of the software under budget and within the expected timeframe in an agile software development environment partly depends on the ability of the development team to predict the number of person-hours (effort) needed by project developers to complete a project. Henceforth, the research may be conducted to investigate the improvement of effort estimation accuracy prediction of agile software development using solo and ensemble techniques with other evaluation metrics. Furthermore, another noteworthy extension is the overall understanding and investigation of how effort estimation in homogeneous or heterogeneous ensemble-based models using the combinations of algorithmic, expert estimation and machine learning techniques can be improved. In effort estimation, the involvement of human experts could include in ensembles. The human expert estimates may maximize the use of context-specific knowledge which may not be accounted for by predictive algorithms, particularly when the development team transitions to working with new emerging technologies or working in new application domains. The improvement of effort estimation accuracy prediction of software development effort by using combinations of estimation techniques involving expert estimation to make ensemble models or frameworks would be another research direction. In the future, we aim to use the knowledge of this investigation and will propose an ensemble model to improve estimation accuracy prediction of software development effort incorporated with algorithmic, expert estimation and machine learning techniques. The proposed model produced at the end of this research will be used by software development firms and practitioners as an instrument to estimate the effort required to develop new software projects at an earlier stage.

**Author Contributions:** Conceptualization, methodology, writing—original draft preparation Y.M.; validation, A.S.K. and M.A.; formal analysis, N.K; investigation, M.A.; resources, funding acquisition, project administration A.A.; review and editing, Y.M, N.K and A.S.K.; supervision, N.K and A.A.; All authors have read and agreed to the published version of the manuscript.

**Funding:** This research is fully funded by Universiti Teknologi Malaysia under the UTM Fundamental Research Grant (UTMFR) with Cost Center No Q.K130000.2556.21H14.

**Acknowledgment:** The authors would like to thank and acknowledge every individual who has been a source of information, support and encouragement on successful completion of this manuscript.

**Conflicts of Interest:** The authors declare no conflict of interest.

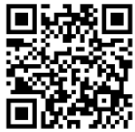
**YASIR MAHMOOD**
https://orcid.org/0000-0003-1578-5229

## Appendix: Primary Studies

S1. Nassif, A. B., et al. (2012). "A Treeboost Model for Software Effort Estimation Based on Use Case Points." 2012 11th International Conference on Machine Learning and Applications: 314-319.

S2. Nassif, A. B., et al. (2011). "Estimating Software Effort Based on Use Case Point Model Using Sugeno Fuzzy Inference System." 2011 IEEE 23rd International Conference on Tools with Artificial Intelligence: 393-398.

S3. Satapathy, S. M. et al. (2016). "Early stage software effort estimation using random forest technique based on use case points." IET Soft 10(1)10-17.

S4. Dan, Z. (2013). "Improving the accuracy in software effort estimation: Using artificial neural network model based on particle swarm optimization." Proceedings of 2013 IEEE International Conference on Service Operations and Logistics, and Informatics.

S5. Malhotra, R. and A. Jain (2011). "Software effort prediction using statistical and machine learning methods." International Journal of Advanced Computer Science and Applications 2(1): 145-152.

S6. Baskeles, B., et al. (2007). "Software effort estimation using machine learning methods." 2007 22nd international symposium on computer and information sciences, IEEE.

S7. Yurdakurban, V. and N. ErdoĞan (2018). "Comparison of machine learning methods for software project effort estimation." 2018 26th Signal Processing and Communications Applications Conference (SIU).

S8. Sharma, P. and J. Singh (2018). "Machine Learning Based Effort Estimation Using Standardization." 2018 International Conference on Computing, Power and Communication Technologies (GUCON).

S9. Idri, A., et al. (2016). "Accuracy comparison of analogy-based software development effort estimation techniques." International Journal of Intelligent Systems 31(2): 128-152.

S10. Rijwani, P. and S. Jain (2016). "Enhanced software effort estimation using multi layered feed forward artificial neural network technique." Procedia Computer Science 89: 307-312.

S11 Nassif, A. B., Capretz, L. F., and Ho, D. (2012). "Estimating software effort using an ANN model based on use case points." Paper presented at the 2012 11th International Conference on Machine Learning and Applications.

S12 Gabrani, G., & Saini, N. (2016). "Effort estimation models using evolutionary learning algorithms for software development." Paper presented at the 2016 Symposium on Colossal Data Analysis and Networking, CDAN 2016, doi:10.1109/CDAN.2016.7570916. (2016).

S13 Burgess, C. J., & Lefley, M. (2001). "Can genetic programming improve software effort estimation? A comparative evaluation." Information and Software Technology, 43(14), 863–873. doi:10.1016/s0950-5849(01)00192-6

S14 S. M. Satapathy, B. P. Acharya, and S. K. Rath, "Class point approach for software effort estimation using stochastic gradient boosting technique," ACM SIGSOFT Softw. Eng. Notes, vol. 39, no. 3, pp. 1–6, 2014.

S15. Jodpimai, P., et al. (2018). "Ensemble effort estimation using selection and genetic algorithms." International Journal of Computer Applications in Technology 58(1): 17-28.

S16. Hosni, M., et al. (2016). "Heterogeneous ensembles for software development effort estimation." Soft Computing & Machine Intelligence (ISCMI), 2016 3rd International Conference on, IEEE.

S17. Kumar, A. and U. Datta (2015). "An Effort Estimation Model for Software Development using Ensemble Learning." International Journal of Computer Applications 115(21).

S18. Wu, D., et al. (2013). "Linear combination of multiple case-based reasoning with optimized weight for software effort estimation." The Journal of Supercomputing 64(3): 898-918.